# Generative AI for Discovering Porous Oxide Materials for Next-Generation Energy Storage

Joy Datta[1], Amruth Nadimpally[2], Nikhil Koratkar[3], Dibakar Datta[1,*]

[1]Department of Mechanical and Industrial Engineering, New Jersey Institute of Technology (NJIT), Newark, NJ 07102, USA

[2] Senior at John P. Stevens High School, Edison, NJ 08820, USA

[3]Department of Mechanical, Aerospace, and Nuclear Engineering, Rensselaer Polytechnic Institute (RPI), Troy, NY 12180, USA

Corresponding Author: Dibakar Datta (dibakar.datta@njit.edu)

**Abstract**

Lithium-ion batteries (LIBs) are essential for modern applications. However, the scarcity of lithium (Li) raises concerns regarding their long-term sustainability. As a result, there is significant interest in exploring alternate battery systems utilizing earth-abundant, low-cost multivalent metals such as aluminum (Al) and calcium (Ca). The primary challenge, however, lies in identifying suitable intercalation hosts for these multivalent-ion batteries. Open-tunnel oxides, characterized by their integrated one-dimensional channels or nanopores, show great promise for facilitating effective ion transport. Yet, the wide range of compositional possibilities makes traditional experimental approaches and quantum-based models impractical for large-scale investigation. In this work, we present a generative artificial intelligence (AI) framework that employs the Crystal Diffusion Variational Autoencoder (CDVAE) and a fine-tuned Large Language Model (LLM) to accelerate the discovery of stable open-tunneled oxide materials for multivalent-ion batteries. We integrate machine learning with data mining techniques to generate five highly promising transition metal oxide (TMO) structures, known for their ability to form open-tunnel oxide frameworks, which we structurally validate using Density Functional Theory (DFT). The generated structures exhibit lower formation energies compared to analogous compositions in the Materials Project (MP) database, suggesting enhanced thermodynamic stability. We further employed the graph-based model M3GNet to relax additional generated structures, providing a computationally efficient alternative to DFT. The selection process is further refined through machine learning-based predictions of formation energy, band gap, and energy above the hull, allowing us to identify materials with significant potential for practical battery applications. This work demonstrates the efficacy of generative AI in efficiently exploring the vast chemical space of transition metal oxides (TMOs), offering a transformative approach to rapidly identify stable, open-tunneled oxides suitable for multivalent-ion batteries. Consequently, it contributes to a more sustainable future for energy storage technologies.

# 1. INTRODUCTION

With the growing need for high performance and sustainable energy storage systems, the development of innovative battery materials has become increasingly crucial[1–3]. Rechargeable batteries, especially those utilizing multivalent ions such as aluminum, calcium, magnesium, and zinc, are recognized as promising substitutes for traditional lithium-ion batteries. This transition is driven by the rising costs of rare-earth lithium and the necessity to identify plentiful, economically viable, and ecologically sustainable alternatives. However, the creation of practical multivalent-ion batteries poses considerable challenges, particularly in identifying host materials that can endure the mechanical strains caused by ion insertion while maintaining robust stability and conductivity[4–7].

The efficacy of battery electrodes is significantly influenced by the selection of particle size[8,9]. While microparticles are preferred due to their cost-effectiveness and increased volumetric energy density[10], they are constrained by mechanical, thermodynamic, and kinetic limitations[7,11,12]. In contrast, nanoparticles provide improved cycle stability, faster charging, and shorter diffusion lengths; however, they exhibit low first-cycle efficiency, reduced volumetric capacity, and higher production costs[13–15]. Multiscale particles (MPs) offer a promising compromise, combining the advantages of both micro- and nanostructures. Niobium tungsten oxide (NTO) and Molybdenum vanadium oxide (MoVO) are examples of naturally porous, open-tunnel oxides that exhibit exceptional ion diffusion due to their nanoscale channels. Although these two materials exist naturally, other multiscale particles, particularly those based on transition metal oxide (TMOs), may further enhance battery performance by leveraging their unique multiscale properties[16].

In this context, TMO-based materials are particularly important because of their distinctive structural features, including rapid ion transport through open channels and exceptional surface area-to-volume ratios[16]. Given the vast range of potential TMO structures from various elemental combinations and stoichiometries, experimental exploration is infeasible, presenting a classic 'needle in a haystack' problem. Moreover, traditional computational techniques such as Density Functional Theory (DFT), although precise, are often too computationally intensive to be employed on a large scale[17,18], especially when dealing with the complex structures of TMO[19]. Therefore, a comprehensive understanding and utilization of training data on TMO-based oxide materials are crucial for discovering and exploring new open-tunnel oxide materials to accommodate ions with multiple charges[20]. This effort must encompass the examination of TMO families based on binary, ternary, quaternary, quinary, and even senary configurations.

In recent years, advanced Machine Learning (ML) techniques have significantly enhanced simulation speed, predictive accuracy, and the discovery of new compounds[21–23]. By leveraging data from Density Functional Theory (DFT) calculations and experimental techniques, ML models are trained using large databases such as the Materials Project Database (MPD)[24], Automatic Flow for Materials Discovery (AFLOW)[25], Inorganic Crystal Structure Database (ICSD)[26] and Open Quantum Materials Database (OQMD)[27]. While forward ML models are effective in predicting material properties and streamlining the screening process[22,28,29], they encounter difficulties to generate entirely new materials, particularly multiscale particles that combine nanoscale and microscale properties[30]. These multiscale structures are critical for next-generation battery technologies, underscoring the limitations of current ML approaches in innovating materials for future needs[31–34].

Recently, models such as CrystalGAN[35], iMatGen[32], and ZeoGAN[36] have employed advanced machine learning techniques, including autoencoders and Generative Adversarial Networks (GANs), to generate distinct crystal structures. Generative models function by constructing a continuous latent space that encodes material data and generates new materials by mapping this latent space to desired properties, offering a potential solution to the long-standing challenge of inverse design. However, accurately translating these latent representations into real 3D structures remains challenging, limiting the practical applicability of certain generative models[30]. Additional issues, such as user-defined post processing, memory-intensive representations, and a lack of invariance to translation and rotation, also pose obstacles. Despite these challenges, the use of generative AI models has emerged as a potent tool in addressing these challenges by significantly expediting the discovery process[30]. Specifically, generative AI models have demonstrated substantial potential in material discovery by facilitating the creation of novel materials with predetermined characteristics identified by the user[37–39]. Through the utilization of generative AI, scientists can efficiently create and assess new materials, therefore simplifying the process of discovery and surpassing the constraints of conventional trial-and-error approaches.

Among these generative models, The Crystal Diffusion Variational Autoencoder (CDVAE)[37] and fine-tuned Large Language Model (LLM)[40–43] represent unique approaches to material discovery. By learning the fundamental distribution of known stable materials, CDVAE can generate a wide range of realistic materials with the required periodic, rotational, and translational symmetries, excelling at constructing stable crystal structures. LLMs, on the other hand, can analyze and comprehend textual data, such as Crystallographic Information Files (CIFs), to produce supplementary structures with specific characteristics, such as porosity, tunnel structure, and conductivity[41,42]. By integrating the advantages of CDVAE and LLM, researchers can intensively investigate potential TMO materials, which are critical for next-generation battery technologies.

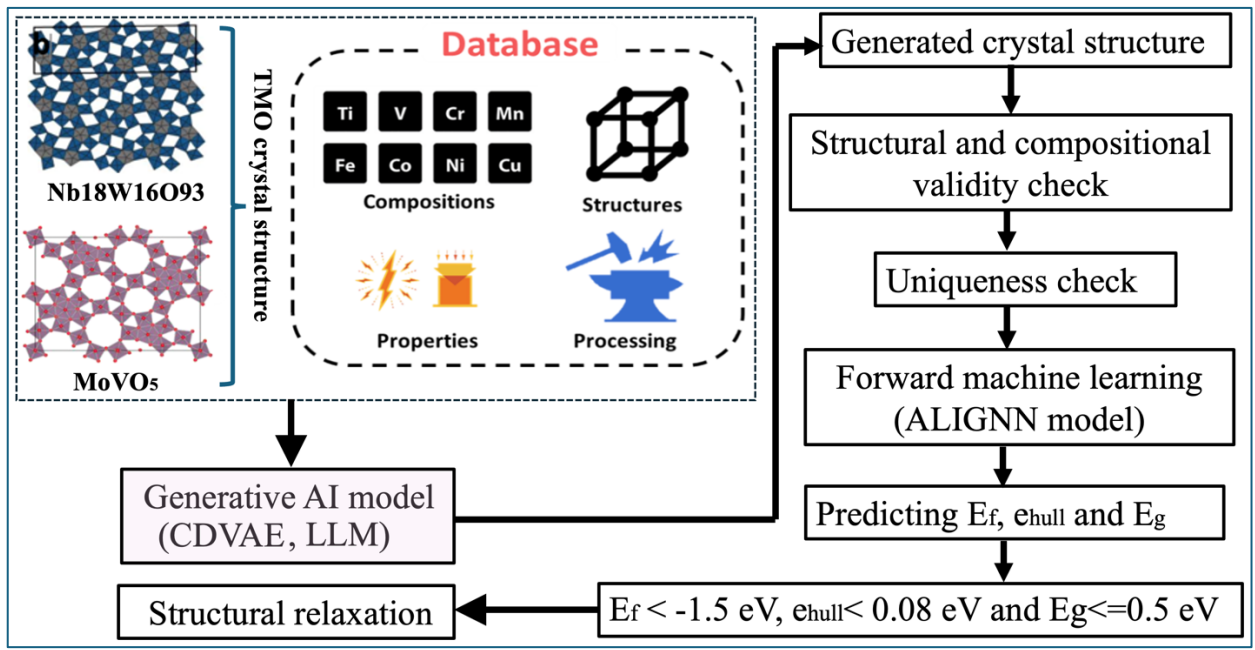

**Figure 1:** An overview of generative AI models, such as CDVAE and a fine-tuned LLM trained on an extensive dataset from the MP database, is presented in this approach for constructing crystal structures. After verifying structural validity and ensuring uniqueness, a forward machine learning model called ALIGNN is used to predict the formation energy, energy above the hull, and band gap of these structures. The thermodynamic and electronic properties are further optimized using machine learning-based structural relaxation methods and DFT calculations to refine the structures.

Our contribution to this work leverages two powerful generative AI models, CDVAE and LLM, to effectively address the challenge of discovering novel TMO based crystal structures. By combining the strengths of CDVAE and LLM, this dual strategy ensures a comprehensive exploration of potential TMO materials. CDVAE generates a broad range of plausible crystal structures, while LLM refines these structures by adding specific, desirable properties. After generating these structures, a forward machine learning model, such as the Atomistic Graph Neural Network (ALIGNN)[44], is employed to predict key properties, such as formation energy, energy above the hull, and band gap. These properties are critical for assessing the thermodynamic and electronic stability of the generated materials, helping to narrow down the selection to the most promising candidates for applications in multivalent-ion batteries. The selected structures then undergo structural relaxation by Density Functional Theory (DFT) to refine their electronic structure and stability. Figure 1 illustrates the workflow. The synergy between AI models and traditional material optimization techniques ensures that the materials align with theoretical predictions and are suitable for real-world applications in advanced battery systems.

## 2. METHODS

### 2.1 Dataset Preparation:

The dataset comprises 44,411 inorganic structures based on TMO materials, including binary, ternary, quaternary, quinary, and senary configurations. Robust methods were employed to analyze the structural traits of these TMO. The distribution of various TMO compositions is illustrated in Fig. 2a.

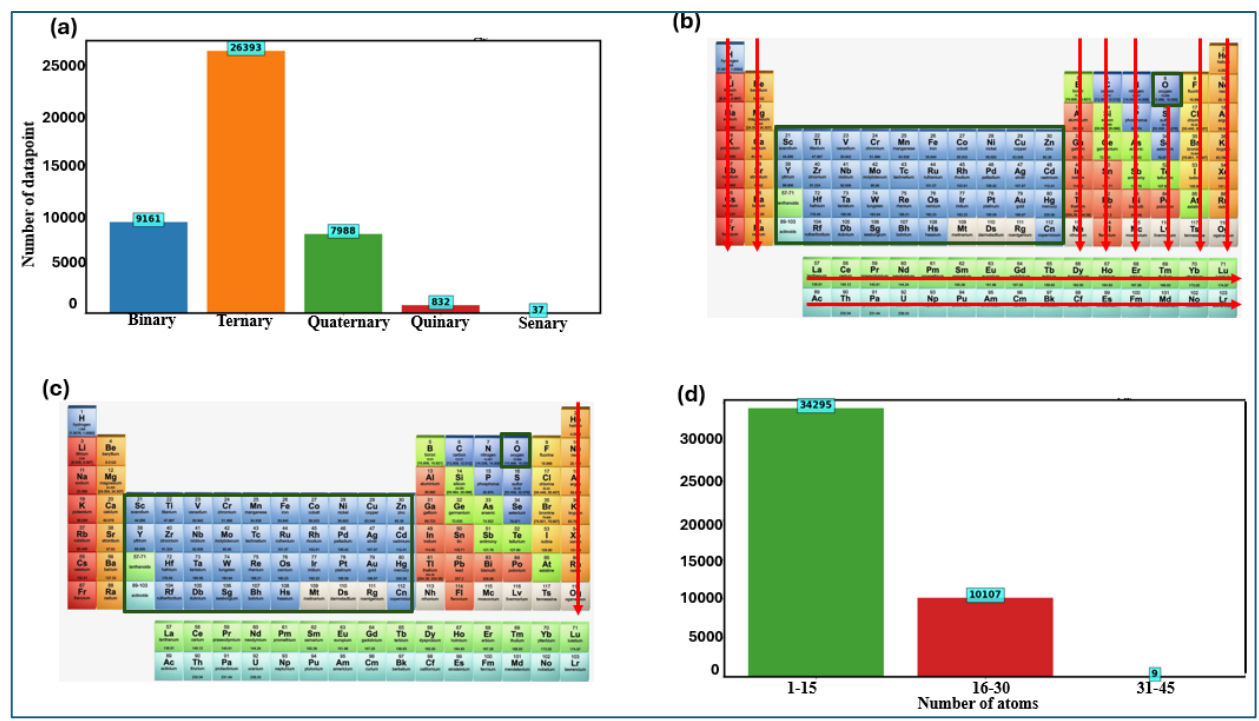

**Figure 2:** Compositional and Atomic Distribution of TMO: (a) Data distribution by TMO composition types, (b) a detailed overview of binary TMOs, (c) an overview of ternary, quaternary, quinary, and senary TMOs, and (d) the distribution of the number of atoms presents in the collected MP dataset. In panels (b)

and (c), materials included in the study are highlighted in green, while excluded materials are shown in red. The uncolored areas in (c) represent the third, fourth, fifth, and sixth components in ternary, quaternary, quinary, and senary TMOs, respectively. An example of a quaternary TMO is represented by the formula $W_sA_rB_mO_n$, where W and A represent any metal from the periodic table, and B is a transition metal. The letters s, r, m, and n denote the stoichiometric ratios. For the ML model, 60% of the data was allocated for training, 20% for testing, and the remaining 20% for validation.

Due to the importance of transition metals in battery materials research, our study focused on these elements. Notably, ternary TMOs constitute approximately 26,393 data points, while senary TMO are underrepresented, with only 37 entries (Fig. 2a). The data were obtained from the Materials Project (MP)[45] database through the API using the Python libraries Matminer[46] and Pymatgen[47]. Fig. 2d shows the distribution of the number of atoms per composition, with a maximum of fewer than 45 atoms considered in the analysis. The entire dataset was divided into three partitions for our ML model: 60% for training, 20% for testing, and the remaining 20% for validation.

## 2.2 Crystal Diffusion Variational Autoencoder (CDVAE)

The CDVAE generative model uses an empirical distribution of known stable materials to learn and produce stable material structures by combining the functionalities of Variational Autoencoders (VAE) and Diffusion models[37]. The model's architecture comprises three primary components: an encoder, an attribute predictor, and a decoder. The encoder, $PGNN_{ENC}$, employs a SE(3)-equivariant periodic graph neural network (PGNN) to transform input data comprising atomic types, atomic coordinates, and a periodic lattice into a lower-dimensional latent space vector z. Maintaining periodic boundaries and respecting symmetries are crucial for accurately representing materials in the latent space[48–50].

Once the latent vector z is generated, the attribute predictor estimates three key properties of the material: composition (c), lattice (L), and the number of atoms (N). These predictions are made using multilayer perceptrons (MLPs), each utilizing a different loss function tailored to the property being predicted. Cross-entropy loss is applied for composition prediction, ensuring the model correctly predicts the ratios of different atom types. For lattice prediction, the model reduces the lattice to six unique, rotation-invariant parameters using the Niggli algorithm[51], which ensures the lattice structure is accurately captured and represented. The L2 loss function is used to minimize the error in this prediction. For the number of atoms, softmax activation is applied, allowing the model to output a probability distribution over a set of possible atom counts. The decoder, $PGNN_{DEC}$, employs Langevin dynamics to reconstruct the material structure from the noisy latent representation. The denoising process iteratively refines atomic types and coordinates to ensure the local and global stability of the created material. The decoder is designed to be SE(3)-equivariant, preserving the material's periodic, rotational, and translational symmetries of the material throughout the denoising process[37].

During training in the VAE framework, the Kullback-Leibler (KL) Divergence Loss ($L_{KL}$) ensures that the latent space z adheres to a normal Gaussian distribution, which is essential for the generative process. This smooth interpolation in the latent space allows for efficient sampling of new materials. The KL divergence loss regularizes the latent space by minimizing the difference between the learned latent distribution q(z|M) and a standard Gaussian distribution p(z), ensuring the model generalizes well for material generation.

Throughout training, the CDVAE model minimizes three primary loss functions: the Aggregated Property Loss ($L_{AGG}$), which evaluates the accuracy of predictions for composition, lattice, and atom count; the Decoder Denoising Loss ($L_{DEC}$), which ensures the decoder accurately reconstructs stable material structures; and the KL Divergence Loss ($L_{KL}$), which regularizes the latent space. After training, the model can select a latent vector z, predict the material's aggregated properties, and initialize an arbitrary material structure to generate new materials. The model applies Langevin dynamics to iteratively refine the structure by adjusting atom types and their positions until a stable material is formed.

A significant innovation in CDVAE is the incorporation of a harmonic force field into the learned gradient field[48]. This integration introduces a physically meaningful inductive bias, allowing for a better approximation of quantum mechanical forces. This process ensures the generation of stable and realistic materials. By employing a well-regularized latent space, noise conditioning, and respect for physical symmetries, CDVAE generates stable, diverse, and accurate materials, making it a powerful tool for material discovery[37,52], particularly in applications like battery materials. The architecture of CDVAE model is shown Fig. 3.

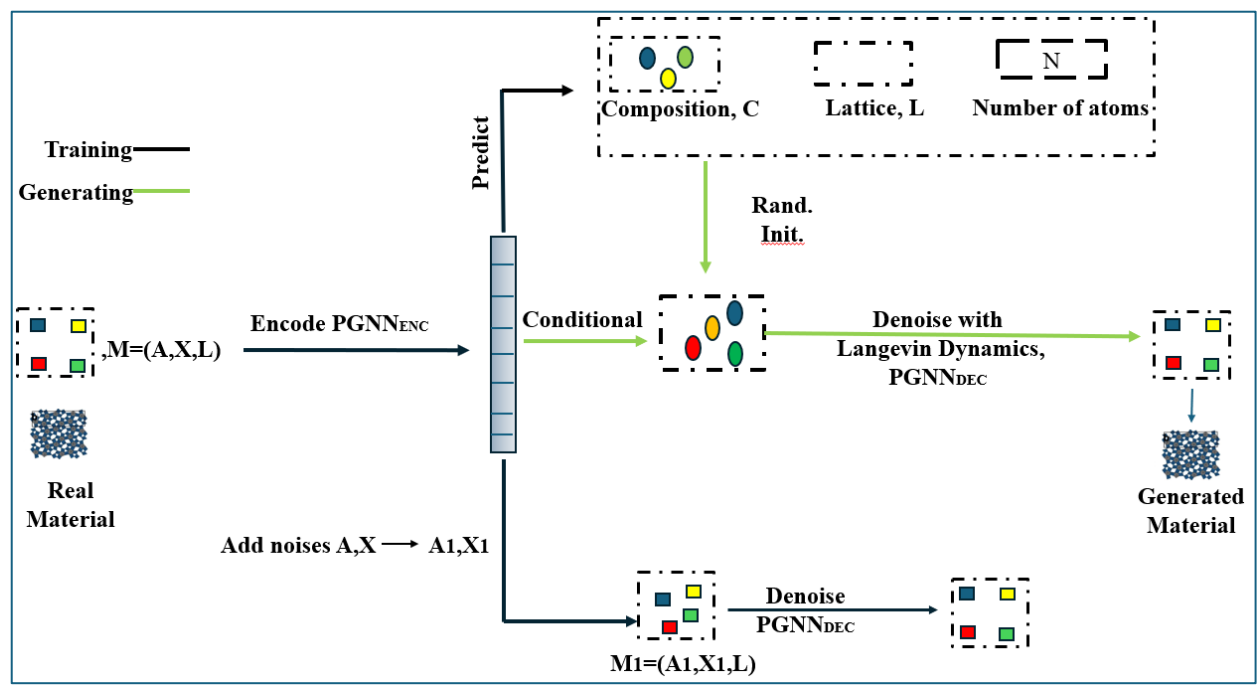

**Figure 3:** Architecture of the CDVAE Model for Generating Stable Material Structures: The model combines a SE(3)-equivariant periodic graph neural network (PGNN) encoder, an attribute predictor, and an SE(3)-equivariant decoder using Langevin dynamics. The encoder maps atomic types, coordinates, and lattice information into a latent space, from which the attribute predictor estimates composition, lattice parameters, and atom count. The decoder refines noisy latent representations through iterative denoising to generate physically stable materials, ensuring periodic and symmetrical properties are maintained.

## 2.3 Fine tuning large language model

The architecture of the Large Language Model (LLM) used in crystal material generation leverages the pre-trained LLaMA-3.1 (8B parameter) model with Unsloth for fine-tuning[40]. Unsloth accelerates the fine-tuning process by 2.1x and reduces memory usage by 60% compared to alternatives such as Flash Attention

2 (FA2)[53] paired with Hugging Face (HF)[54]. This fine-tuning strategy improves training efficiency while lowering computational costs.

The process begins with transforming crystal structures into a format compatible with the language model, represented as a tuple of lattice vectors, angles, atomic identities, and coordinates. This data is then tokenized into sequences, with each digit of the numerical values tokenized separately. The pre-trained LLaMA-3.1 model, originally designed for natural language tasks, is adapted for atomic structure prediction using LoRA (Low Rank Adaptation)[55], reducing the need to retrain the entire model while maintaining high accuracy. The model can handle various generation tasks, including unconditional synthesis of new crystal structures, text-conditional generation based on material properties, and infilling to modify existing structures. To capture symmetries in crystal structures, data augmentations are applied during training, ensuring that the model learns translation and permutation invariance without requiring special architectural modifications[43,56]. Hyperparameters such as temperature and nucleus size are adjusted during generation process to balance randomness and stability in the resulting crystal structures. Our approach of leveraging the LLM model through fine-tuning is illustrated in Fig. 4.

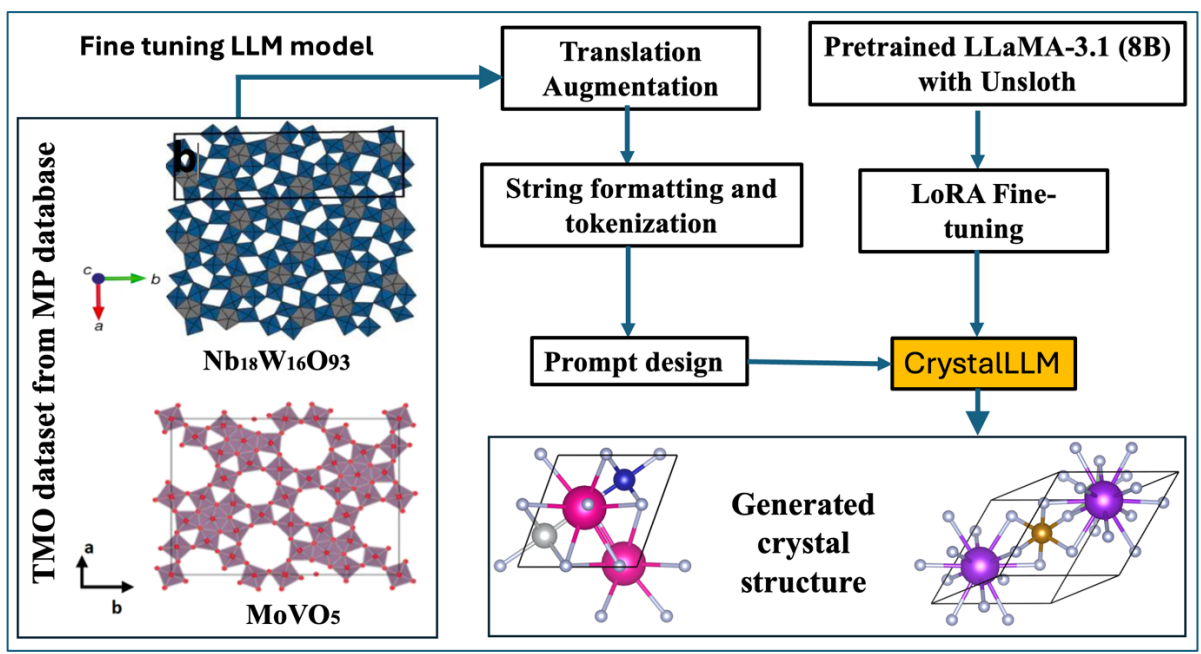

**Figure 4:** Overview of fine-tuning LLaMA 3.1 (8B) model for generating crystal structures. LoRA involves training only a subset of the model parameters (i.e., the low-rank matrices) to adapt the pre-trained model to crystal generation task.

### 2.4 Atomistic graph neural network (ALIGNN):

For our forward ML, we choose the ALIGNN model[44,57]. By fusing bond and line graph structures, the ALIGNN model enhances atomic and bond representations through a unique graph convolution approach. Bond graphs, similar to traditional atomistic models, represent atoms as nodes and bonds as edges, while line graphs represent bonds as nodes and connect pairs of bonds that share a common atom. This distinctive arrangement improves the aggregation of bond-specific data, allowing for more accurate updates to atomic and bond level representations.

To incorporate angular information, ALIGNN constructs the line graph from the bond graph by transforming bonds into nodes and defining edge characteristics using radial basis function (RBF) expansions of bond angle cosine values. With the inclusion of this angular data, the model can better predict material properties by capturing finer details of atomic structures. The model's primary goal is to accurately represent the complex bond-atom interactions in crystalline materials. By efficiently transferring information between the bond and line graphs, ALIGNN improves the prediction of material properties, utilizing edge-gated graph convolution and angular information to effectively capture key structural features that influence the behavior of crystalline materials[58].

### 2.5 M3GNet model:

M3GNet is a graph-based deep learning model that utilizes Interatomic Potential (IAP) to provide a highly effective and precise alternative to Density Functional Theory (DFT) in predicting relaxed energies of materials[59]. It employs Graph Neural Network (GNN) to represent materials as mathematical graphs, with atoms serving as nodes and the connections between them as edges. The architecture of M3GNet effectively captures many-body interactions and accurately represents the Potential Energy Surface (PES). Through iterative graph convolution processes, M3GNet enhances the accuracy of energy, force, and stress predictions by updating bond, atom, and global state information.

In contrast to DFT, which is characterized by high computational costs and reduced scalability with system size, M3GNet offers a faster, linear-scaling approach without compromising accuracy[52]. The M3GNet model, trained on a comprehensive dataset from the MP database, has strong generalization capabilities across a diverse array of chemistries and crystal structures. Its demonstrated ability to efficiently relax millions of theoretical structures makes it a highly suitable tool for large-scale material exploration. In this context, the relaxed energies of 55 produced structures were computed using M3GNet, offering a cost-effective alternative to DFT for efficient energy reduction and reliable material identification.

### 2.6 DFT Relaxation:

For the relaxation of generated selected crystal structures, we integrated the Projector Augmented Wave (PAW) approach into the Vienna Ab initio Simulation Software (VASP) and conducted structural optimizations using DFT. The Perdew-Burke-Ernzerhof (PBE) variant of the Generalized Gradient Approximation (GGA) has been performed to model the exchange-correlation interactions. We employed varying energy cutoff ranging from 200 eV to 500 eV, depending on the specific crystal structure being optimized. These calculations were based on the plane wave basis set. Brillouin zone sampling utilized a different k-point grid depending on the selected crystal structures' structural parameters. We employed an energy convergence criterion of $10^{-6}$ eV/Å and a force convergence tolerance of 0.02 eV/Å to relax all the selected structures.

The adsorption potential (V) is calculated as:

$$V = \frac{\Delta G}{n_f} \tag{1}$$

where $n_f$ is the concentration of single atom. The Gibbs free energy, $\Delta G$, is defined as:

$$\Delta G = \Delta E_f + P\Delta V_f - T\Delta S_f \tag{2}$$

In equation (2), at room temperature, $P\Delta V_f$ and $T\Delta S_f$ are negligible compared to $\Delta E_f$ [8]. The formation energy, $\Delta E_f$, can be computed using the equation:

$$\Delta E_f = \Delta E_{total} - (\sum n_i E_i) \tag{3}$$

where $E_{total}$ is the Gibbs free energy of the generated TMO-based crystal structure. $n_i$, is the number of atoms of element $i$ in the TMO structure. $E_i$ is the Gibbs free energy of the elemental form of atom $i$. A negative value of $\Delta E_f$ indicates that the compound is thermodynamically stable relative to its constituent elements.

## 3. RESULTS AND DISCUSSION

The CDVAE model is configured for efficient performance, using a latent dimension of 256 and hidden layers between 128 and 256 dimensions. It employs a GemNet-dQ[60] as the decoder, with a 7.0 Å radius and a maximum of 20 neighbors per atom. Langevin dynamics[48] refines noisy structures by improving atomic positions and bond preferences with diffusion parameters such as $\sigma_{begin} = 10.0$ and $\sigma_{end} = 0.01$.

The loss function is optimized using weighted costs for several model characteristics, including coordination (cost coord=10), atom types (cost type=1), and lattice stability (cost lattice=10). High-cost values, such as 10 for coordination and lattice stability, indicate that the model emphasizes these features for precise structure formation during training. To balance convergence speed and stability, the learning rate is initially set to 0.001 with a minimum value of 0.0001, preventing overshooting and stalling during training by allowing efficient parameter updates.

The Niggli-reduced lattice[51] ensures unique lattice prediction, while CrystalNN[61] and KNN algorithms capture periodicity and atom-atom interactions. The model is SE(3) invariant, with DimeNet++[62] serving as the encoder, ensuring robustness against rotational, translational, and periodic boundary conditions. We utilize an L40 architecture with 48 GB VRAM, 250 GB RAM, and 16 vCPUs for training the CDVAE model on our customized CIF dataset. For our CDVAE model, training for 5000 epochs took approximately 60 hours, demonstrating its computationally efficiency and speed.

The CDVAE model generated 10000 structures, which were subjected to a series of precise screening and validation steps to ensure they met the necessary standards. The first step involved verifying that each structure satisfied important performance criteria, considering both structural soundness and compositional balance.

For structural verification, we applied the method suggested by Court et al.[63], which checks the distance between atom pairs to ensure no atoms are closer than 0.5 Å. This method focuses solely on the distance between atoms, ignoring other factors. For composition, we used the SMACT method to verify charge neutrality. If the SMACT data indicates that the material has an overall neutral charge, it is considered compositionally correct. We then filtered out generated crystal structures already present in the MP database by checking for similar compositions. After applying these filters for structural and compositional validity, and ensuring uniqueness, 8203 out of the 10000 structures passed the initial screening.

Next, we aimed to identify materials with desirable properties using a forward ML approach named ALIGNN to predict the formation energy/atom, band gap, and energy above hull. For our forward ML model, we have predicted all the property with pretrained ALIGNN model. We selected a formation energy baseline of -1.5 eV/atom, as lower values indicate higher synthesizability and stability for battery materials, following Ren et al.[38] criteria for energy applications. For band gap, the lower band gap has high probability to utilize into battery applications due to its high conductivity[64,65]. Therefore, we selected structures with a band gap less than or equal to 0.5 eV/atom. Lastly, the most important property of ensuring stability is to choose the energy above hull. Kim et. al[30] considered less than 0.08 eV/atom energy above hull as a standard margin to consider any generated material to be stable and synthesizable. Therefore, our study follows the same approach.

After applying these property-based filters, we retrieved 42 structures from the CDVAE approach, guided by their potential to expand battery material discovery. The selection included 5 oxygen-containing structures and 37 oxygen-free structures. Of these, 21 structures matched existing entries in the MP database but offered new configurations with differences in stoichiometry, lattice parameters, or space groups, The remaining 21 structures were entirely novel.

Despite being oxygen-free, the 37 structures were generated using TMO-trained models, inheriting key beneficial features associated with energy storage applications. These oxygen-free structures demonstrate notable compositional and structural diversity, which could give rise to unique ionic diffusion pathways, improved electronic conductivity, and increased structural stability - key properties typically associated with TMO-based battery materials.

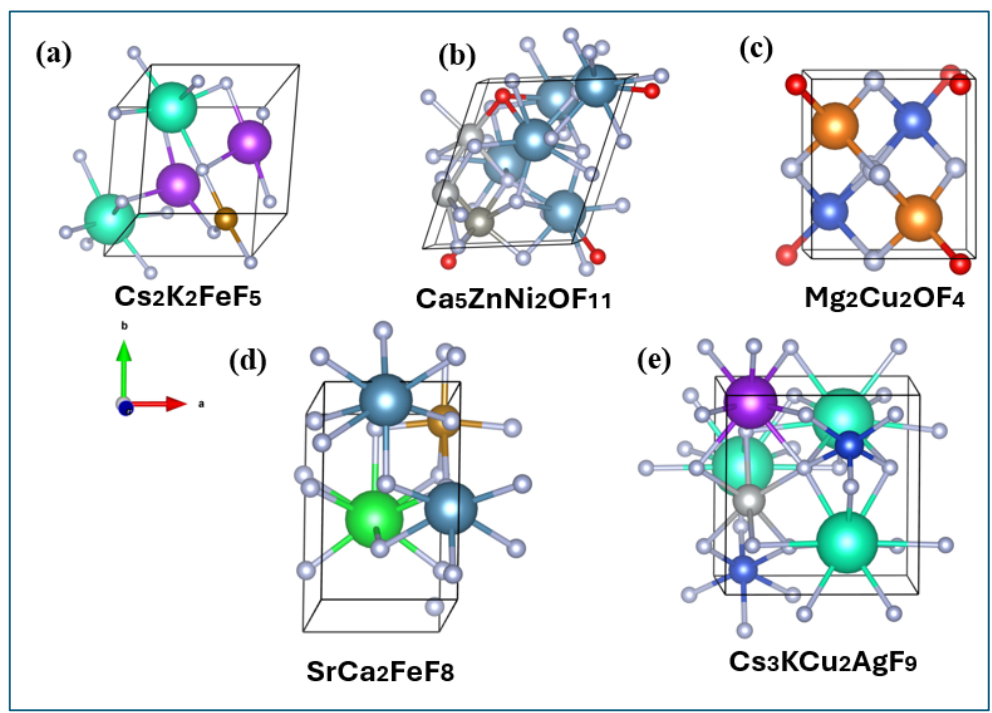

**Figure 5:** Overview of the top 5 most stable structures generated by the CDVAE model, ranked by energy above hull (eV/atom).

It is essential to recognize that non-oxide materials, particularly those containing transition metals, may offer valuable advantages due to their ability to accommodate multiple ions while maintaining good conductivity[66,67]. Given that the generated structures retain critical characteristics of TMO-based materials, they are strong candidates for further investigation. Predictions using the ALIGNN model indicate that these 42 structures satisfy essential stability and performance benchmarks, underscoring the value of both oxygen-containing and oxygen-free candidates in advancing battery technologies.

This approach not only capitalizes on the established strengths of TMO systems but also opens up promising new avenues for material discovery and innovation in the development of next-generation energy storage solutions. Out of these 42 structures, we present the 5 most stable structures, as predicted by the ALIGNN model based on energy above the hull, shown in Fig. 5.

For fine-tuning the LLM model, this study calibrates the Meta-Llama-3.1-8B model using a parameter-efficient method specifically tailored for generating crystal structures. The Low-Rank Adaptation (LoRA) technique is employed to minimize computational burden without compromising performance. The algorithm is configured with a rank of 8, an alpha of 32, and a dropout rate of 0.05. These parameters are selected to achieve optimal model performance on hardware with limited resources. Additionally, 4-bit quantization is used to reduce memory consumption while maintaining the precision of the process.

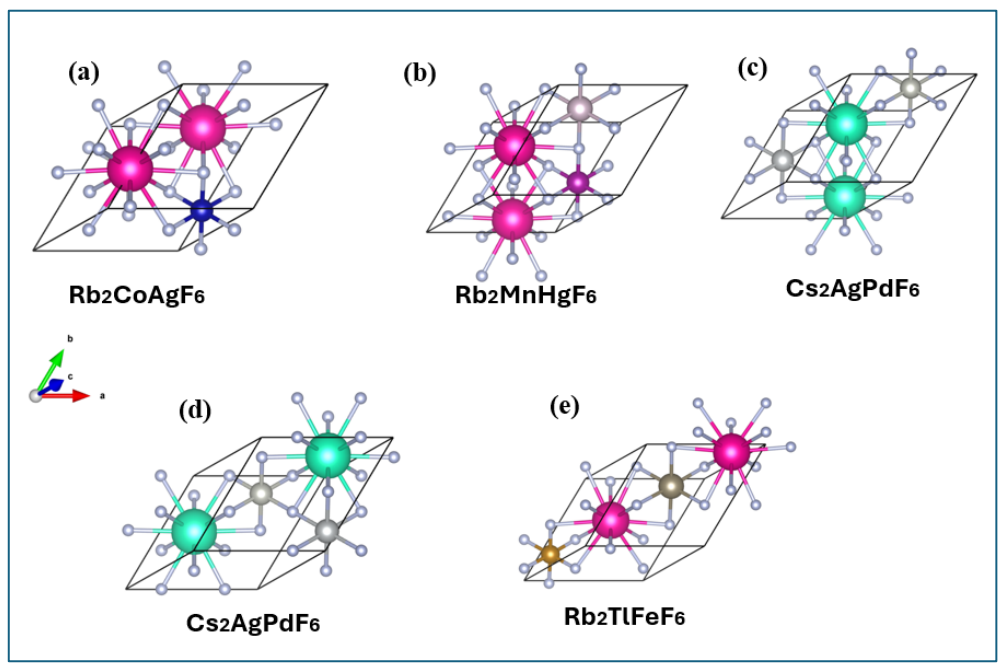

**Figure 6:** Overview of the top 5 most stable structures generated by the LLM model, ranked by energy above hull (eV/atom).

Material parameters such as formation energy, band gap, and energy above hull are represented by tokenized sequences derived from CIF. The model performs tasks such as generating entire structures, property-based conditional generation, and infilling missing components. To accommodate the model's size and complexity, training is conducted with a batch size of 1 and gradient accumulation of 1. A learning rate of $1\times10^{-4}$, along with a cosine scheduler and 100 warmup steps, is used to ensure stability. The model is trained for 25 epochs, with periodic checkpoints and assessments carried out during the process. We utilized the same GPU architecture for training the fine-tuned LLM model. Despite the model being quantized, the training process took 150 hours, demonstrating that even with optimizations, fine-tuning large language models remains computationally expensive.

Our fine-tuned LLM model generated 10000 structures according to the specified instructions. After applying compositional, structural validity, and uniqueness checks, 1087 structures remained. We then employed a similar forward machine learning approach to filter these structures based on the three properties mentioned earlier. Following this filtration, only 18 structures passed the criteria.

In the case of the LLM model, all 18 final structures are oxygen-free, consistent with the rationale of the CDVAE model discussed earlier. Despite the absence of oxygen, these structures retain key properties typically associated with TMO due to the model being trained entirely on TMO data. Out of these 18 structures, 16 are present in the MP database but show differences in elemental ratios, crystal dimensions, or symmetry classifications, while the remaining 2 structures are entirely novel, presenting new opportunities for material design. The five most stable structures, selected based on their lowest energy above the hull, are presented in Fig. 6.

Fig. 7 illustrates the distribution of key material properties, including formation energy, band gap, and energy above the hull, for both the CDVAE and LLM models. For the CDVAE model, the formation energy (Fig. 7a) ranges from approximately -3.83 eV/atom to -1.73 eV/atom, indicating that the model produces more energetically stable structures. The most stable structure has a formation energy of -3.83 eV/atom, suggesting that the CDVAE model can generate highly stable materials. The band gap distribution (Fig. 7b) covers a range of -0.09 eV to 0.49 eV, indicating moderate variability in the electronic properties of the generated structures. The energy above the hull (Fig. 7c) ranges from -0.07 eV/atom to 0.08 eV/atom, signifying the relative stability of the structures, with the potential for generating both stable and metastable structures.

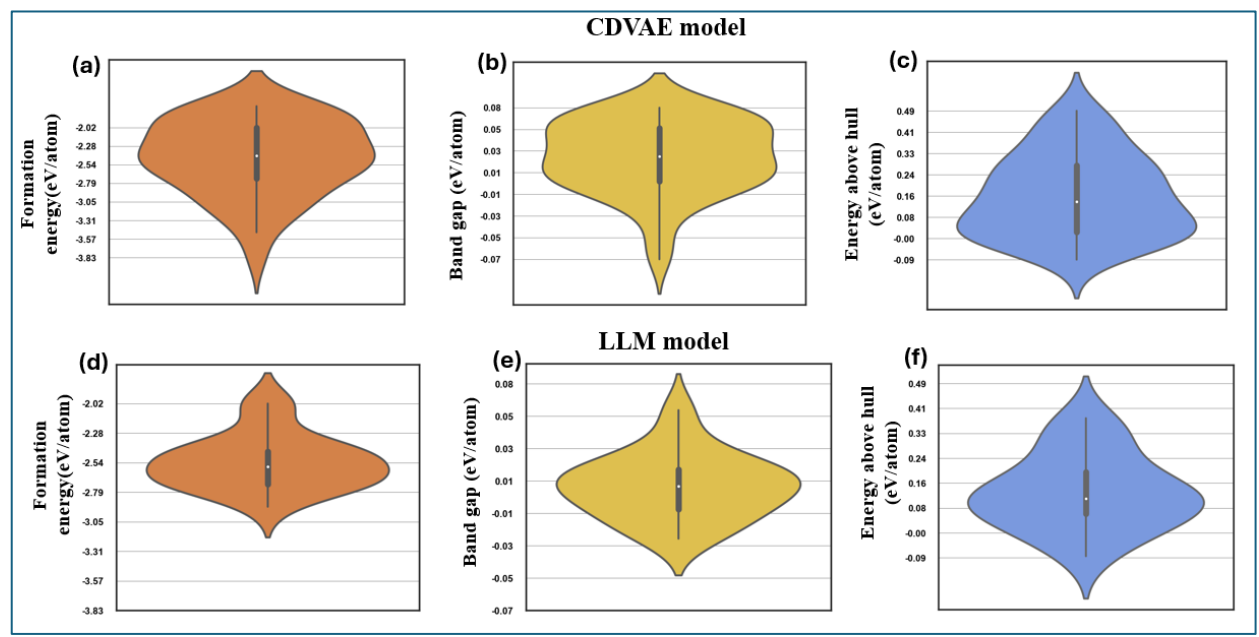

**Figure 7:** Property distribution of filtered structures before relaxation: Data distribution of formation energy (eV/atom), band gap (eV), and energy above hull (eV/atom) for the (a-c) CDVAE and (d-f) LLM model. The CDVAE model generates more diverse and stable structures compared to the LLM model, with the most stable structure having a formation energy of -3.85 eV/atom.

In comparison, the LLM model (Fig. 7d-f) displays a narrower distribution of formation energy, ranging from -2.92 eV/atom to -2.02 eV/atom, with fewer energetically stable structures compared to the CDVAE model. The band gap distribution (Fig. 7e) spans from -0.08 eV to 0.08 eV, reflecting limited diversity in electronic properties. The energy above the hull for the LLM model (Fig. 7f) ranges from -0.03 eV/atom to 0.06 eV/atom. While this is comparable to the CDVAE model, it suggests that the LLM model generates fewer highly stable structures overall.

These results demonstrate that the CDVAE model outperforms the LLM model in terms of stability, as reflected by its lower formation energy and energy above the hull. This highlights its ability to generate more stable and diverse structures compared to the LLM model, which exhibits a narrower range of property distributions and produces less stable configurations.

For structural relaxation, the M3GNet model was applied to all 42 filtered structures from the CDVAE model, while only 13 out of 18 filtered structures from the LLM model successfully optimized. This highlights the CDVAE model's ability to generate more stable structures. Fig. 8 compares the relaxed energy distributions of both models. The CDVAE model (Fig. 8a) produces a wider range of relaxed energies, with several structures achieving lower energy values down to -6.28 eV/atom, indicating greater stability. In contrast, the LLM model (Fig. 8b) shows a narrower distribution, with the most stable structure having a relaxed energy of around -4.92 eV/atom. These findings confirm the CDVAE model's superiority in generating more diverse and stable structures after relaxation.

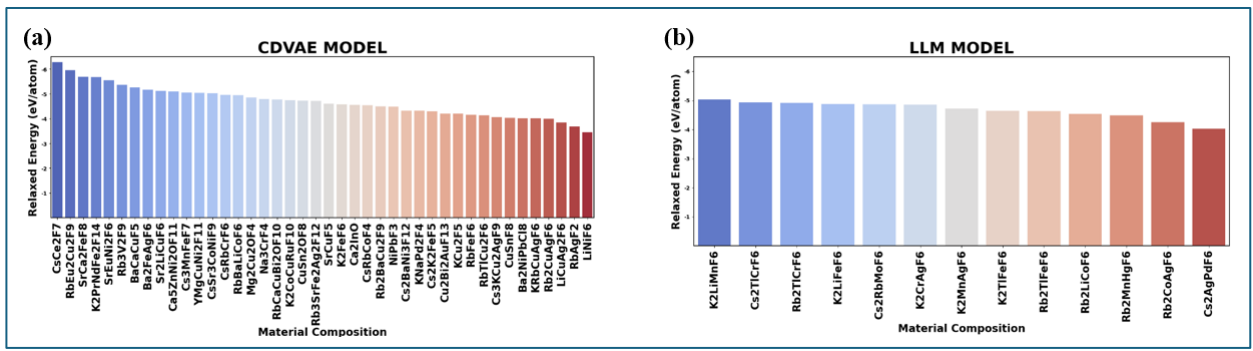

**Figure 8:** Relaxed energy distribution of generated structures using the M3GNet model: (a) 42 crystal structures generated by the CDVAE model, and (b) 13 crystal structures generated by the LLM model.

The focus on TMO-based structures in our study is driven by their critical role in advancing multiscale battery technologies. While our models, particularly the LLM, generated a diverse range of materials, none of the LLM-generated structures met the criteria for TMO-based compositions, highlighting a gap in that specific material space. In contrast, the CDVAE model successfully produced five TMO-based structures, which are of particular interest due to the well-documented advantages of TMOs in energy storage applications, such as their ability to enhance ionic conductivity, structural stability, and multivalent ion accommodation.

Given the importance of TMO materials for high-performance multivalent-ion batteries, it is essential to validate and thoroughly analyze the TMO structures generated by the CDVAE model. By performing DFT calculations, we ensured the structural and thermodynamic stability of these materials, confirming their suitability for real-world applications. All five structures successfully relaxed during the DFT process. Next, we calculated the formation energies for these five compositions using equation (3). The required individual atomic energies for the calculation were also determined using DFT. Fig. 9 compares the formation energies per atom between our generated structures and the corresponding structures from the MP database, where available. These five TMO-based structures (Fig.10), feature large open-tunnel frameworks designed to enhance ion transport by accommodating multivalent ions. Their interconnected channels make them promising candidates for high-performance multivalent-ion batteries, offering both structural stability and efficient ion diffusion pathways.

We confirm that three out of the five compositions are present in the MP database, albeit with different stoichiometry ratios. For comparison, we selected the most stable composition from the MP database based on their energy above the hull values and extracted their formation energies. As shown in Fig. 9, the generated structures demonstrate superior stability, as reflected by their lower formation energy per atom compared to the MP database counterparts. For instance, the formation energy of generated $Mg_2Cu_2O_4F_4$ is about -1.8 eV/atom, while the corresponding MP database structure has a formation energy of -1.6 eV/atom. Similarly, for $Ca_2OIn$, the generated structure exhibits a significantly lower formation energy of -6.5 eV/atom compared to -3.0 eV/atom for the MP database equivalent. These comparisons further illustrate the enhanced stability of the generated structures relative to their MP database counterparts.

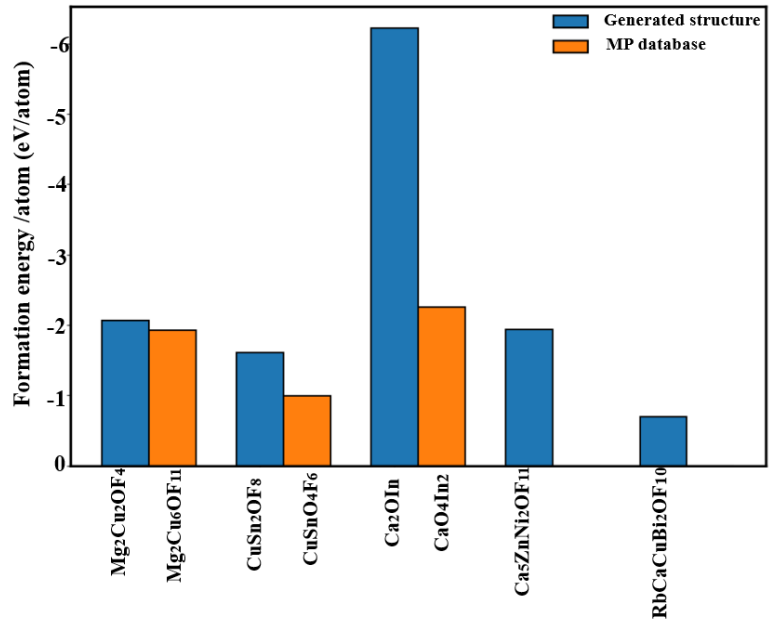

**Figure 9:** Comparison of formation energy per atom for generated TMO structures versus MP database entries. The generated structures exhibit lower formation energies, indicating higher stability compared to their MP database counterparts.

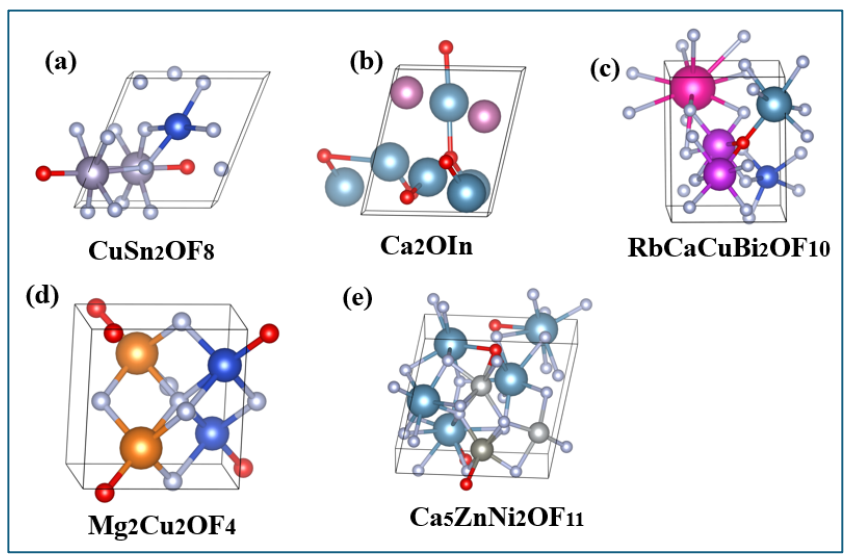

**Figure 10:** Overview of five TMO-based structures generated by the CDVAE model, highlighting large open-tunnel frameworks designed for efficient accommodation of multivalent ions in battery applications.

## 4. CONCLUSION

In this study, we demonstrate the efficacy of combining generative AI models—CDVAE and fine-tuned LLM—to address the challenges of discovering novel TMO-based materials for multiscale batteries. Five of the generated structures satisfy the requirements for open-tunnel oxide frameworks, which are essential for efficient ion transport in multivalent-ion battery systems. Due to their interconnected one-dimensional channels and nanopores, these TMO-based structures show great promise as battery materials.

The CDVAE model, which specializes in generating stable crystal structures, successfully created five TMO compositions, all of which were structurally confirmed using DFT relaxation. Our comparison with the MP database reveals that the generated structures have lower formation energies than their MP counterparts, indicating higher stability. For instance, the structure of $Ca_2OIn$ exhibits a much lower formation energy (-6.5 eV/atom) compared to the MP database's -3.0 eV/atom.

We find that the CDVAE model outperforms the LLM in terms of structural stability and variety. One reason the LLM model underperforms is that it was trained on a 4-bit quantized model, which may have limited its capacity to generate highly stable structures. We anticipate that increasing the quantization to an 8-bit model could enhance its performance and enable the generation of more stable structures. Our future direction includes exploring other open-source models to discover additional TMO-based materials.

Our findings underscore the potential of generative AI in accelerating the identification of stable, open-tunnel TMO materials, which are essential for the development of next-generation multivalent-ion batteries. The combination of AI-driven models and traditional computational approaches offers a transformative method for material discovery, significantly reducing the time and computational costs typically associated with trial-and-error experiments. This framework presents a novel approach for discovering stable and efficient materials for sustainable energy storage devices.

## AUTHOR INFORMATION

### Corresponding Author

Dibakar Datta
Email: dibakar.datta@njit.edu ; Phone: +1 973 596 3647

### Author Contributions

J.D., D.D. and N.K. conceived the project. J.D. performed all work and wrote the manuscript with D.D. and N.K.  A. N. helped J.D. with calculations. All authors approved the final version of the manuscript.

## CONFLICT OF INTEREST STATEMENT

The authors have no conflicts of interest to declare. All authors have seen and agree with the contents of the manuscript and there is no financial interest to report. We certify that the submission is original work and is not under review at any other publication.

## ACKNOWLEDGEMENT

The work is supported by National Science Foundation (NSF) (Award Number #2237990). Authors acknowledge Advanced Cyberinfrastructure Coordination Ecosystem: Service & Support (ACCESS) for the computational facilities (Award Number - DMR180013).

## DATA AVAILABILITY

Data can be obtained by requesting the corresponding author.

## CODE AVAILABILITY

Code for extracting the data and its training model is available on GitHub (https://github.com/joy1303125/Generative-AI-for-battery-material), providing researchers in the field with a valuable resource. Restrictions apply to the availability of the simulation codes, which were used under license for this study.

Performance NiS2 Hollow Nanosphere Cathodes in Magnesium-Ion Batteries Enabled by Tunable Redox Chemistry. *Nano Lett.* **2022**, *22* (24), 10184–10191.